\documentclass[twocolumn,superscriptaddress,showpacs,prl,amsmath,amstex,amssymb,citeautoscript,longbibliography]{revtex4-1}
\pdfoutput=1
\usepackage{natbib}
\usepackage[english]{babel}
\usepackage{letltxmacro}
\usepackage{latexsym}
\LetLtxMacro{\ORIGselectlanguage}{\selectlanguage}
\makeatletter
\DeclareRobustCommand{\selectlanguage}[1]{%
  \@ifundefined{alias@\string#1}
    {\ORIGselectlanguage{#1}}
    {\begingroup\edef\x{\endgroup
       \noexpand\ORIGselectlanguage{\@nameuse{alias@#1}}}\x}%
}
\newcommand{\definelanguagealias}[2]{%
  \@namedef{alias@#1}{#2}%
}
\makeatother
\definelanguagealias{en}{english}
\definelanguagealias{English}{english}
\usepackage{graphicx}
\usepackage{amsmath}
\usepackage{amsfonts}
\usepackage{amssymb}
\usepackage{bm}
\usepackage{color}
\usepackage{soul} 
\usepackage{amssymb}
\usepackage{wasysym}
\usepackage{dsfont}
\usepackage{float}

\usepackage{hyperref}
\hypersetup{
    bookmarks=false,         
    unicode=false,          
    pdftoolbar=false,        
    pdfmenubar=true,        
    pdffitwindow=false,     
    pdfstartview={FitH},    
    pdftitle={The effect of $SU(2)$ symmetry on many-body localization and thermalization},    
    pdfauthor={Ivan V. Protopopov, Wen Wei Ho, and Dmitry A. Abanin},     
    pdfsubject={},   
    pdfcreator={},   
    pdfproducer={}, 
    pdfkeywords={many-body localization} {matrix elements} {disordered systems}, 
    pdfnewwindow=true,      
    colorlinks=true,       
    linkcolor=black,          
    citecolor=blue,        
    filecolor=magenta,      
    urlcolor=blue           
}

\newcommand{\be}{\begin{equation}}
\newcommand{\ee}{\end{equation}}
\newcommand{\bea}{\begin{eqnarray}}
\newcommand{\eea}{\end{eqnarray}}

\newcommand{\la}{\langle}
\newcommand{\ra}{\rangle}

\begin{document}
\title{The effect of $SU(2)$ symmetry on many-body localization and thermalization}

\author{Ivan V. Protopopov}
\affiliation{Department of Theoretical Physics, University of Geneva, 24 quai Ernest-Ansermet, 1211 Geneva, Switzerland}
\affiliation{Kavli Institute for Theoretical Physics, University of California, Santa Barbara, CA 93106, USA}
\author{Wen Wei Ho}
\affiliation{Department of Theoretical Physics, University of Geneva, 24 quai Ernest-Ansermet, 1211 Geneva, Switzerland}
\affiliation{Kavli Institute for Theoretical Physics, University of California, Santa Barbara, CA 93106, USA}
\author{Dmitry A. Abanin}
\affiliation{Department of Theoretical Physics, University of Geneva, 24 quai Ernest-Ansermet, 1211 Geneva, Switzerland}
\affiliation{Kavli Institute for Theoretical Physics, University of California, Santa Barbara, CA 93106, USA}

\date{\today}
\begin{abstract}

The many-body localized (MBL) phase is characterized by a complete set of quasi-local integrals of motion and area-law entanglement of excited eigenstates. We study the effect of non-Abelian continuous symmetries on MBL, considering the case of $SU(2)$ symmetric disordered spin chains. The $SU(2)$ symmetry imposes strong constraints on the entanglement structure of the eigenstates, precluding conventional MBL. We construct a fixed-point Hamiltonian, which realizes a non-ergodic (but non-MBL) phase characterized by eigenstates having logarithmic scaling of entanglement with the system size, as well as an incomplete set of quasi-local integrals of motion. We study the response of such a phase to local symmetric perturbations, finding that even weak perturbations induce multi-spin resonances. We conclude that the non-ergodic phase is generally unstable and that $SU(2)$ symmetry implies thermalization. The approach introduced in this work can be used to study dynamics in disordered systems with non-Abelian symmetries, and provides a starting point for searching non-ergodic phases beyond conventional MBL. 

\end{abstract}
\pacs{72.15.Rn,75.10.Pq,05.30.Rt}
\maketitle

{\it Introduction.} Over the past several years, the phenomenon of many-body localization (MBL) has been attracting significant interest, both theoretically~\cite{Basko06,Mirlin05,OganesyanHuse,Znidaric08,PalHuse,Vosk13,Serbyn13-2,Serbyn13-1,Huse13,Moore12,Alet14,Demler14,Pekker14,Ponte15,Lazarides15,Abanin20161,Khemani16} and experimentally~\cite{Bloch15,Bloch16,Monroe16}. Many-body localization occurs in strongly disordered systems and is driven by a mechanism similar to the (single-particle) Anderson localization in the many-body Hilbert space. Isolated many-body localized systems exhibit zero conductivity and avoid thermalization, and therefore provide the only known, generic example of ergodicity breaking in many-body systems.

MBL eigenstates have low, area-law entanglement entropy~\cite{Serbyn13-1,Bauer13}, in contrast to the excited eigenstates of ergodic systems, which have thermal, volume-law entanglement. The systems in which all states are many-body localized exhibit a new kind of robust integrability: a complete set of quasi-local integrals of motion (LIOMs) emerges~\cite{Serbyn13-1,Huse13} (see also~\cite{Imbrie16,ScardicchioLIOM,Chandran14}). 
Apart from providing a simple physical intuition for the ergodicity breaking in MBL phase, LIOM theory has been used to explain dynamical properties of MBL eigenstates, including logarithmic entanglement growth in a quantum quench setup~\cite{Serbyn13-2,Serbyn13-1,Huse13}, as well as power-law decay~\cite{Serbyn14}  and revivals~\cite{Vasseur14} of local observables, which can be tested in cold atoms experiments. 
 
A natural question concerns the role of various symmetries on MBL and thermalization. 
Previous works focused mostly on MBL in the presence of discrete symmetries, such as $\mathbb{Z}_2$ symmetry. It was shown~\cite{HuseSondhi13,Pekker14,Kjall14} that in this case two distinct MBL phases are possible, one of which locally preserves $\mathbb{Z}_2$ symmetry, while the other phase locally breaks that symmetry. It was also argued that MBL can protect topological~\cite{HuseSondhi13,Bauer13} and symmetry-protected topological~\cite{Chandran15} order at finite temperatures. Ref.~\cite{Vasseur15} considered the effect of a particular non-Abelian discrete symmetry on MBL. 

The goal of this paper is to study disordered systems with {\it continuous non-Abelian symmetries}. We focus on the simplest and experimentally  relevant example of such a symmetry -- $SU(2)$ spin rotation symmetry -- which is realized in the random Heisenberg spin-$1/2$ chain: 
\be\label{eq:hamiltonian}
H=\sum_{i=1}^{L} J_i {\bf s}_i\cdot {\bf s}_{i+1}, 
\ee
where coupling $J_i$ are randomly drawn from some distribution, and ${\bf s}_i=(s_i^x,s_i^y,s_i^z)$ are the Pauli operators. 

$SU(2)$ symmetry puts severe constraints on the entanglement structure of the eigenstates and on the locality properties of the integrals of motion in a possible non-ergodic phase. Generally, in the presence of $SU(2)$ symmetry, it is impossible to have eigenstates with area-law entanglement, and therefore conventional MBL cannot occur~\cite{Chandran15,Vasseur16}. As we will argue below, the symmetry does allow eigenstates with entanglement that grows logarithmically with the system size. Simultaneously, at least some integrals of motion must become non-local. The key question  then is whether such a non-ergodic (but non-MBL) phase may be stable. Below, we will perform the stability analysis, finding that in general such an entanglement structure is unstable, in the sense that in a sufficiently large system, an arbitrarily weak perturbation of the Hamiltonian inevitably strongly mixes eigenstates, leading to delocalization. We will discuss the delocalization mechanism, finding that, although inevitable, thermalization can be parametrically slow at strong disorder. Our analysis therefore indicates that $SU(2)$ symmetry is inconsistent with non-ergodicity, and implies thermalization. We expect that the approach introduced below can be used in future studies of ergodicity breaking beyond conventional MBL. 

We note that two recent works~\cite{VasseurHotChains,Demler15} used real-space strong disorder renormalization group (RSRG) for excited states to analyze the behaviour of disordered, $SU(2)$-symmetric spin chains. Our approach allows us to take into account multi-spin processes, which are not captured by RSRG. As we will see below, such processes inevitably lead to delocalization. The relation between the present work and RSRG approach is discussed at the end of the paper.


{\it Conventional MBL phase.} Let us start by recalling the description of the conventional MBL phase (no non-Abelian symmetries) in terms of LIOMs. The defining property of MBL is that highly excited eigenstates can be obtained from non-entangled product states by a quasi-local unitary transformation $U$, $U^\dagger H U=H_{\rm diag}$. For the case of spin-$1/2$ chains (such as random-field XXZ model that has been extensively studied~\cite{PalHuse,Moore12,Serbyn13-2,Alet14,Serbyn15}), it is convenient to choose a product state basis in which eigenstates have definite $s_i^z$ projections. Then, operators $\tau_i^z=Us_i^z U^\dagger$, which are dressed spin operators, are quasi-local integrals of motion. In terms of these operators spins, the Hamiltonian takes a simple form~\cite{Huse13,Serbyn13-1}: 
\be\label{eq:fixed_point_MBL}
H=\sum_i h_i \tau_i^z +\sum_{i,j} J_{ij} \tau_i^z \tau_j^z + \sum_{i,j,k} J_{ijk} \tau_i^z \tau_j^z \tau_k^z +\dots, 
\ee
where couplings $J_{ij..k}$ decay exponentially with distance. The Hamiltonian (\ref{eq:fixed_point_MBL}) is often viewed as a ``fixed-point'' Hamiltonian of the MBL phase. Importantly, integrability is robust: if a weak perturbation which is a sum of local, but otherwise arbitrary terms is added to it, a new set of quasi-local integrals of motion can be defined. In what follows, we will show that the presence of $SU(2)$ symmetry significantly modifies the possible structures of integrals of motion and of the fixed-point Hamiltonian. 

\begin{figure}[t]
\includegraphics[width=1\columnwidth]{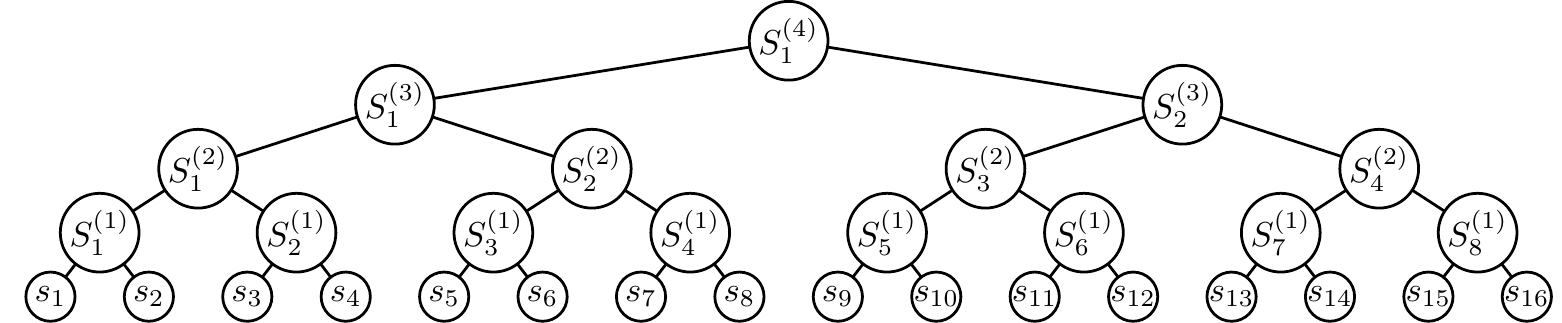}
\caption{\small The simplest possible tree-structure and corresponding  LIOMs. }
\label{Fig:SimpleTree}
\end{figure}

{\it SU(2) symmetry implies non-local integrals of motion.} Let us now discuss the possible structure of the non-ergodic phase in an $SU(2)$ symmetric spin chain (\ref{eq:hamiltonian}). First, we argue that the eigenstates cannot be area-law entangled. Consider two initially disconnected systems, $\mathcal L$ and $\mathcal R$. The symmetry dictates that each eigenstate $|\alpha\ra_{\mathcal L}$ of $\mathcal L$ belongs to a multiplet $m_{\mathcal L}$ with some total spin $S_{\mathcal L}$ and degeneracy $2S_{\mathcal L}+1$. Similarly, an eigenstate $|\beta\ra_{\mathcal R}$ of $\mathcal R$ belongs to a $(2S_{\mathcal R}+1)$-degenerate multiplet $m_{\mathcal R}$ with spin $S_{\mathcal R}$. 
When we couple the two systems, even a very weak coupling will force the eigenstates of the $\mathcal{L}+\mathcal{R}$ system to transform as an irreducible representation of the SU(2) symmetry acting on the combined system. The least entangled (and therefore most non-ergodic) states correspond to the scenario when different multiplets (which are eigenstates of disconnected ${\mathcal L}$ and ${\mathcal R}$ systems) do not strongly hybridize once the systems are joined. 
Assuming that this holds, the eigenstates of the whole system are obtained by adding together a single multiplet $m_{\mathcal L}$ with spin $S_{\mathcal L}$ and a single multiplet $m_{\mathcal R}$ with spin $S_{\mathcal L}$ to to form larger multiplets, whose spin can take values $|S_{\mathcal L}-S_{\mathcal R}|, |S_{\mathcal L}-S_{\mathcal R}|+1,\dots, S_{\mathcal L}+S_{\mathcal R}$. Such states have entanglement entropy which is typically of the order $S_{\rm ent}\sim\ln ({\rm min} (S_{\mathcal R}, S_{\mathcal L}))$. Since the total spin of each subsystem grows extensively with its size, we conclude that if a non-ergodic phase exists in the presence of $SU(2)$ symmetry, the eigenstates cannot be area-law entangled, as in the MBL phase.

 \begin{figure}[t]
 \includegraphics[width=0.8\columnwidth]{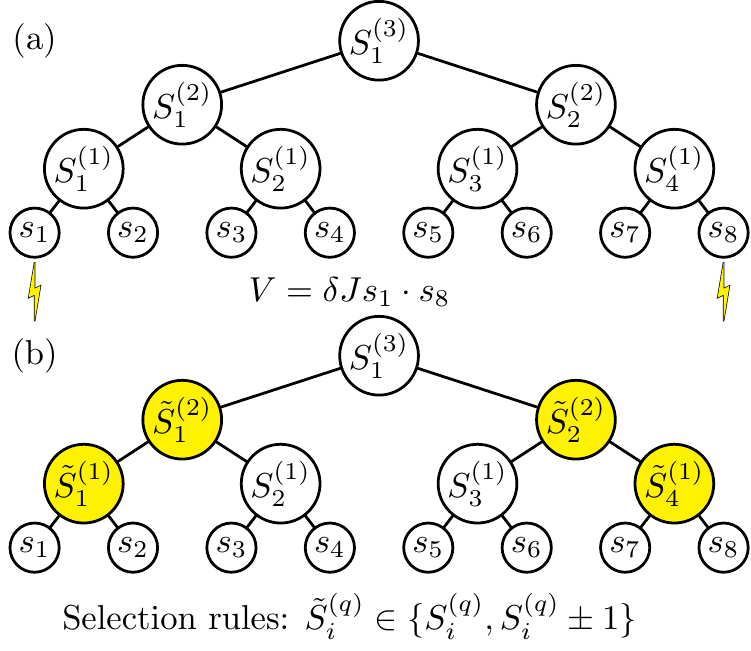}
 \caption{Illustration of selection rules for the matrix elements of the operator~${\bf s}_1\cdot {\bf s}_L$ for a system of $8$ spins. The two states (upper (a) and lower (b) trees) have non-vanishing matrix element only when $\tilde{S}=S,\, S\pm 1$ while spins in the other nodes coincide.  }
 \label{Fig:SelectionRules}
 \end{figure}

Further, as the system size is increased, spins of different subsystems (``blocks") have to be added up, such that larger and larger spins are formed~\cite{Chandran15}. 
Depending on the strength of the coupling between different spins (which are disordered) a spin of a given block should be first added with the spin of the block to the left or to the right of it. Graphically, we can denote adding two spins by connecting corresponding blocks; then, a tree-like structure, an example of which is shown in Fig.~\ref{Fig:SimpleTree} emerges. An eigenstate is uniquely specified by the spin values at every leaf of the tree. Such a plausible structure of the eigenstates, described by a tree tensor network, corresponds to logarithmic scaling of entanglement entropy with the system size, $S_{\rm ent}\sim \ln L$. Such entanglement scaling is non-MBL, but also strongly sub-thermal, and therefore describes non-ergodic eigenstates.



The picture described above corresponds to an incomplete set of quasi-local integrals of motion: a total spin at every step is approximately conserved; it becomes a precise IOM if we deform it by a quasi-local unitary operator, which accounts for perturbative mixing between different multiplets. It is instructive to write down a complete set of integrals of motion for the case of  a regular tree structure, illustrated in Fig.~\ref{Fig:SimpleTree}. There, at the first step, spins $2i-1$ and $2i$ are added to form a (possibly larger) spin ${\bf S}_i^1$, then at the second step spins ${\bf S}_{2i-1}^{(1)}$, ${\bf S}_{2i}^{(1)}$  are added to form spin ${\bf S}_i^{(2)}$, etc., until we get just one large spin describing a multiplet of the whole system. In this case, the complete set of IOMs is given by:
\begin{equation}
\label{eq:IOMs}
 \left[{\bf S}_i^{(1)}\right]^2, \,\, i=1,..\frac{L}{2}; \;\; \left[{\bf S}_i^{({2})}\right]^2, \,\, i=1,...\frac{L}{4},...
\end{equation}
where $L$ is the total number of spins $1/2$ in the chain. The IOMs $ \left[{\bf S}_i^{(k)}\right]^2$ become less and less local as $k$ is increased: they act on $2^k$ spin-$1/2$s. This should be contrasted with the conventional MBL phase characterized by a complete set of quasi-local IOMs. 

We note that since the order in which blocks should be merged depends on the value of the total spin of the resulting block, the integrals of motion would have a different structure for different states. However, for strong disorder such an ambiguity in the block merging only arises when the spin of the resulting block is chosen to be small. As the block merging progresses, such a situation becomes less and less probable. We can therefore assume that, at least starting from some high enough level of our hierarchical construction,  the order of merging is fixed and does not depend on the state.

{\it Fixed-point Hamiltonian.}  It is possible to construct a quasi-local ``fixed-point'' Hamiltonian, for which operators (\ref{eq:IOMs}) are exact IOMs (note that there is a whole family of such fixed-point Hamiltonians): 
\begin{widetext}
\be\label{eq:fixed-point}
H_{\text{FP}}=\sum_{i=1}^{L/2} J_i^{(0)} {\bf s}_{2i-1} \cdot {\bf s}_{2i}+\sum_{i=1}^{L/4} J_{i}^{(1)} {\bf S}_{2i-1}^{(1)} \cdot {\bf S}_{2i}^{(1)}+\sum_{i=1}^{L/8} J_{i}^{(2)} {\bf S}_{2i-1}^{(2)} \cdot {\bf S}_{2i}^{(2)}+\dots
\ee
\end{widetext}
where the couplings $J_i^{(a)}$ are random, and in order for the Hamiltonian to be quasi-local, they should decay exponentially,  $J_i^{(k)}\propto 2^{-\gamma k}$ with $\gamma>2$~\footnote{Spin ${\bf S}_i^{(k)}$ is obtained by adding $2^k$ spins $1/2$, and is less or equal than $2^{k-1}$. 
Thus, the magnitude of each interaction term at level $k$ is $J_{i}^k {\bf S}_{2i-1}^{(k)} \cdot {\bf S}_{2i}^{(k)} \sim 2^{-\gamma k} 2^{2(k-1)}$. 
We see that in order for this term to decay with $k$, condition $\gamma>2$ must hold.
}.

{\it (In)stability of the non-ergodic phase.} The Hamiltonian (\ref{eq:fixed-point}) has eigenstates which are minimally entangled, given the symmetry constraints. The key question is whether such eigenstates are stable with respect to small, but finite, $SU(2)$-symmetric perturbations of the fixed-point Hamiltonian (\ref{eq:fixed-point}). Or, equivalently, whether such a structure of eigenstates can naturally arise starting from a generic local Hamiltonian (\ref{eq:hamiltonian}). 

To answer this question, we study the stability of the eigenstates described by a tree tensor network with respect to local perturbations of the Hamiltonian, thereby extending the approach of Ref.~\cite{Serbyn15}. 
We choose an $SU(2)$-symmetric perturbation which couples the ends of the spin chain:
\be\label{eq:local_pert}
\hat{V}=\delta J \, {\bf s}_1 \cdot {\bf s}_L.
\ee
Such a perturbation describes changing the boundary condition from open to periodic. 
 
\begin{figure}[t]
\includegraphics[width=0.9\columnwidth]{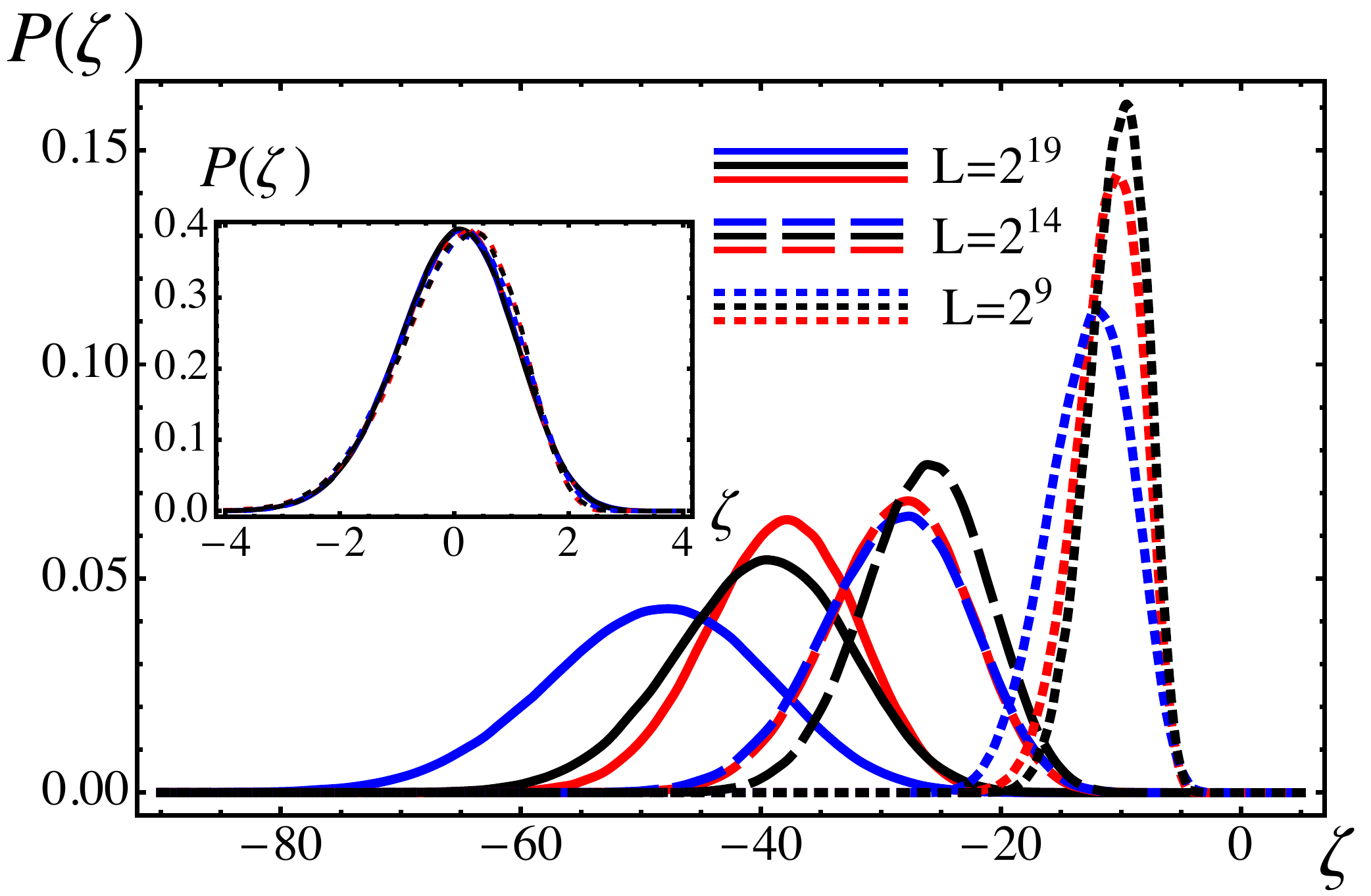}
\caption{Distribution of $\zeta=\log_2 V$ for different sizes of the system: $L=2^9$ (dotted lines), $L=2^{14}$ (dashed lines) and $L=2^{19}$ (full lines). 
For each system size three different base states were randomly generated. The corresponding distributions are shown by different colors.  The inset shows the result of scaling the distributions to the corresponding widths. 
 }
\label{Fig:distr}
\end{figure}
 
The problem of finding the eigenstates of $H_{\text{FP}}+\hat{V}$ can be formulated as a hopping problem on a lattice, where sites are eigenstates $|\alpha\ra$ of $H_{\text{FP}}$. Each state $|\alpha\ra$ is uniquely specified by choosing values of the total spins for each node of the tree, as well as the total $z$-projection of the spin at the last node. Each site $|\alpha\ra$ has on-site energy $E_\alpha=\la\alpha | H_{\text{FP}} |\alpha\ra$, and hopping between sites is set by the matrix elements of the perturbation:
\be\label{eq:hopping}
V_{\alpha\beta}=\la\alpha| V|\beta\ra.
\ee
In order to analyze the statistics of matrix elements, we first note that due to global $SU(2)$ symmetry the perturbation does not change the value of the total spin. Further, the symmetry imposes stringent selection rules on the matrix elements, see Fig.~\ref{Fig:SelectionRules}. The spins on the branches of the tree not involving spin $s_1$ are not affected by operator ${\bf s}_1$. Moreover, it is possible to show~\footnote{In  the language of group theory operators $s_1^{x, y, z}$ organize spin $1$ representation of $SU(2)$. Correspondingly, the matrix elements of ${\bf s}_1$ between two states of the system with total spin $S$ and $S^\prime$ can be non-zero only  if the product of the $S$, $S^\prime$ and spin-1 representations contains trivial representation as a direct summand. This is only the case for $S-S^\prime=0,\, \pm 1$.}  that ${\bf s}_1$ can change any spin on the branch of the tree which involves spin $s_1$  by $\pm 1$, or leave it unchanged. Noting that there are ${\log _2 L}$ nodes in this branch, as well as the fact that $\hat{V}$ can also change the spins on the branch involving ${\bf s}_n$, we obtain that the operator (\ref{eq:local_pert}) couples a given base state to 
$$
K(L)\approx 3^{2\log _2 L}=L^{\alpha}, \qquad \alpha=2\log_2 3
$$
other states~\footnote{In addition to the matrix elements ruled out by the selection rules stated in the text there are other zero $V_{\alpha\beta}$, see also Ref.~\cite{SOM} for more details. However, for a typical state these exceptions do not influence the scaling of connectivity with the system size.}.

Another important ingredient is the level spacing for the hopping problem. We note that in a system with a quasi-local Hamiltonian, a local perturbation can only significantly couple eigenstates with energy difference of order $J$ (where $J$ is the typical interaction scale of the Hamiltonian). Thus, the level spacing in the manifold of states to which a given state is coupled, can be estimated as~\footnote{If the couplings $J_i^k$ decay as $2^{-\gamma k}$ with $\gamma\gg 1$ the in the fixed-point  Hamiltonian (\ref{eq:fixed-point}) dictates   the scaling of the {\it minimal} level spacing $\Delta_{\rm min}\propto L^{-\gamma +1/2}$. For $\gamma<2\log_2 3+1/2$ we however expect the level spacing to obey Eq. (\ref{Eq:Delta})}
\begin{equation}
\Delta(L)\approx \frac{J}{K(L)}.
\label{Eq:Delta}
\end{equation}

{\it Matrix elements.} Matrix elements connecting a given base state to the ones allowed by the selection rules  can be decomposed in terms of Clebsch-Gordan coefficients~\cite{SOM}, and are readily accessible for numerical simulations.  In Fig. \ref{Fig:distr} we show the  distribution of the quantity $\zeta=\log_2 |V/ \delta J|$ for several randomly chosen base states in systems of size $L=2^{9}, 2^{14}$ and $2^{19}$ . We observe that for each of the bases states the distribution of $\zeta$ is well behaved and can be characterized by the mean $\overline{\zeta}$ and standard  deviation $\sigma$ (depending on both the system size and particular base state chosen).  
Moreover, the inset in Fig. \ref{Fig:distr} shows that,  after shifting by respective $\overline{\zeta}$ and scaling by the corresponding $\sigma$, all those distributions collapse into a single, which is approximately Gaussian:
\begin{equation}
P(\zeta)\propto e^{-(\zeta-\bar{\zeta})/2\sigma^2}.
\label{Eq:P}
\end{equation}
The mean $\bar{\zeta}$ and standard deviation $\sigma$ in Eq. (\ref{Eq:P}) are random quantities which depend on the base state. We were able to analyze their statistics in the limit of large $L$.  We computed the average of $\bar{\zeta}$~\cite{SOM}, finding that it grows linearly with $\log L$:
\begin{equation}
\langle\bar{\zeta}\rangle= -\beta \log_2 L+O(1)\,, \qquad \beta =\frac{17+4\ln 2}{9\ln2},
\label{eq:beta}
\end{equation}
while the standard deviation of $\bar{\zeta}$ and the average of $\sigma$  obey
\begin{equation}
\sqrt{\langle(\bar{\zeta}-\langle\bar{\zeta}\rangle)^2\rangle}\propto \sqrt{\log_2 L}\,, \qquad \langle\sigma\rangle=C\sqrt{\log_2 L}.
\label{eq:sigma}
\end{equation}
The constant $C$ is of order one~\cite{SOM}. 

{\it Instability with respect to local perturbation.} Next, we use the scalings of the level spacing and typical matrix element, Eqs.~(\ref{Eq:Delta}) and (\ref{Eq:P}),(\ref{eq:beta}),(\ref{eq:sigma}) to estimate the probability of finding a (long-range) resonance induced by a local perturbation. To that end, we consider a ratio $g=\frac{V}{\Delta(L)}$, which can be viewed as the Thouless parameter. Neglecting the fluctuations in the level spacing, the probability of finding a resonance is given by:
\be\label{eq:prob}
P(g>1)=P\left(  \zeta>A  \right), \,\, A=\log_2 \left(  \frac{J}{\delta J K(L)}\right). 
\ee
Furthermore, for fixed $\bar\zeta$ and $\sigma$, this probability can be transformed as follows:
\be\label{eq:PA}
P\left(  \zeta>A |\bar{\zeta},\sigma \right)=\frac{1}{2} {\rm erfc}\left( \frac{A-\bar\zeta}{\sqrt{2}\sigma} \right), 
\ee
where ${\rm erfc}(x)$ is the complementary error function. Using the above results (\ref{Eq:Delta}),(\ref{eq:beta}), we obtain that the asymptotic behavior of the expression in the r.-h.s. of the above equation is given by: 
\be\label{eq:asymptotic}
 \frac{A-\bar\zeta}{\sqrt{2}\sigma}\propto  \frac{\log_2 \frac{J}{\delta J}-(\alpha-\beta)\log_2 L}{C\sqrt{\log_2 L}},
\ee
where $C$ is the proportionality constant of order 1 in Eq.(\ref{eq:sigma}).
Interestingly, we find that the parameter $\gamma=\alpha-\beta\approx 10^{-4}$ is positive, but extremely small. The above expression approaches zero beyond length scale $L_c$ given by equation 
$$
\log_2 \frac{J}{\delta J}\approx C\sqrt{\log_2 L_c}, 
$$
which yields 
\be\label{eq:Lc}
L_c\approx \left( \frac{J}{\delta J}\right)^{\frac{\log_2(J/\delta J)}{C^2}}. 
\ee
When $L>L_c$, the Thouless parameter becomes $g\approx 1/2$, which signals instability of tree-like eigenstates and delocalization, for arbitrarily small $\delta J$. This implies that the new eigenstates will have volume-law entanglement entropy, indicating that the non-ergodic phase described above is intrinsically unstable. However, since not all states are resonant with its nearest neighbours, we expect that the system will show non-trivial, and possibly glassy dynamics. \footnote{We also note that in the limit $\delta J\to 0$, the delocalization occurs at the lengthscale $L_*\approx \left(\frac{J}{\delta J} \right)^{1/\gamma}$, which corresponds to the numerator in the r.-h.s. of Eq.(\ref{eq:asymptotic}) becoming negative. However, given the smallness of parameter $\gamma$, $L_*<L_c$ only for extremely small $\delta J\lesssim J\cdot 2^{-C^2/\gamma}$, and therefore practically, the delocalization lengthscale will be given by Eq.(\ref{eq:Lc}).} 


{\it Relation to RSRG-X.} Previous studies by Vasseur et al.~\cite{VasseurHotChains} and Agarwal et al.~\cite{Demler15} employed real-space renormalization group procedure (RSRG-X) to analyze the eigenstates of $SU(2)$-symmetric spin chains. In this procedure, strongly coupled pairs of spins are identified, and their spins are added. Such a procedure naturally gives rise to a tree-like structure of the eigenstates described above. We note that Refs.~\cite{VasseurHotChains,Demler15} used different RG rules, and seemingly arrived at opposite conclusions: the  former study found that the RSRG-X procedure breaks down for $SU(2)$ chains, and based on that concluded that thermalization was inevitable, while the latter one claimed that the procedure remains well-defined.

We emphasize that our approach allows us to analyze multi-spin processes. Conventional RSRG, in contrast, only allows one to analyze short-range resonances, which involve a finite number of nearby spins. Our results show that, even if the RSRG-X procedure for $SU(2)$-symmetric systems remains well-defined as one increases the system size, delocalization occurs due to multi-spin processes which are not accounted for by RSRG-X. We therefore expect that a proper analysis of dynamics in disordered $SU(2)$ spin chains should involve a combination of RSRG-X and the approach introduced in this paper. The role of the perturbation introduced by hand above will be played by smaller terms which are neglected in the RSRG-X approach.

{\it Summary.} We have studied the effect of $SU(2)$ symmetry on MBL. We argued that, in contrast to conventional systems where MBL occurs, such symmetry dictates that the eigenstates have larger than area-law entanglement, and some integrals of motion must become non-local. We have introduced a fixed-point Hamiltonian which gives rise to non-ergodic eigenstates with $S_{\rm ent}\sim \ln L$ entanglement. Further, we showed that a weak, local perturbation inevitably introduces resonances between the eigenstates in a sufficiently large system, leading to delocalization. Our results indicate that $SU(2)$ symmetry necessarily implies delocalization; while proving thermalization is a difficult (if not impossible) task, we expect that $SU(2)$-symmetric systems thermalize. 

We expect that our approach can be extended to other systems to analyze the consistency of the RSRG-X procedure. We also expect that the approach described here can be useful in analyzing the effect of other non-Abelian symmetries (both continuous and discrete), and, more generally, in searching for non-ergodic phases that have a richer entanglement structure than the conventional MBL systems with area-law entanglement. 

{\bf Acknowledgements.} We thank Eugene Demler, Maksym Serbyn, and Romain Vasseur for insightful discussions and helpful comments on the manuscript. This research was supported by the Swiss NSF, and in part by the NSF under Grant No.~PHY11-25915. We thank the Kavli Institute for Theoretical Physics at Santa Barbara, where a part of this work was done, for hospitality during the program {\it Synthetic Quantum Matter}.

\bibliography{mbl_SU2}

\widetext
\vspace{2cm}
\begin{center}
\textbf{\large Supplemental Material to ``The effect of $SU(2)$ symmetry on many-body localization and thermalization''}
\end{center}
\setcounter{equation}{0}
\setcounter{figure}{0}
\setcounter{table}{0}
\setcounter{page}{1}
\setcounter{section}{0}
\setcounter{subsection}{0}
\setcounter{subsubsection}{0}
\makeatletter
\newcommand{\mysection}[1]{
 \refstepcounter{section}
  \section*{ {\bf S-\Alph{section}. } #1} }
  
  \newcommand{\mysubsection}[1]{
 \stepcounter{subsection}
  \subsection*{ {\bf S-\Alph{section}\arabic{subsection}. } #1} }
   \newcommand{\mysubsubsection}[1]{
 \stepcounter{subsubsection}
  \subsubsection*{ #1} }
\renewcommand{\thesection}{S-\Alph{section}}

\renewcommand{\theequation}{S-\Alph{section}.\arabic{equation}}
\renewcommand{\thefigure}{S\arabic{figure}}
\renewcommand{\bibnumfmt}[1]{[S#1]}
\renewcommand{\citenumfont}[1]{S#1}

\begin{figure}
\includegraphics[width=500pt]{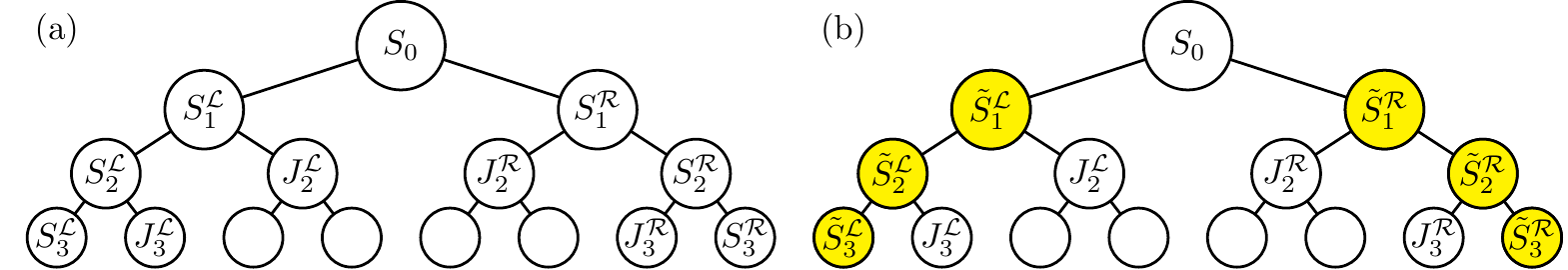}
\caption{Two states of the system connected by an operator ${\bf s}_1\cdot{\bf s}_L $ }
\label{Fig:States}
\end{figure}

\mysection{Introduction}
In this supplemental material  we derive the scaling of a typical matrix element {$V = \langle \alpha | O | \beta \rangle $} of the operator  $O = {\bf s}_1\cdot {\bf s}_L$ between  two tree-like states $| \alpha \rangle, |\beta\rangle$  presented in the main text. Figure~\ref{Fig:States} depicts the scenario considered, and we elaborate on the notation used below.

\mysubsection{Notations}

 We consider two states of a length $L=2^{{\cal {\cal K}}-1}$ spin chain given by two trees of depth ${\cal {\cal K}}$, and will henceforth use the term `state' and `tree' interchangeably.  
We denote by $S_i^{\mathcal L}$ and $S_i^{\mathcal R}$, $i=1, \ldots {\cal {\cal K}}-1, $ the spins in the left-most and right-most branches of one tree, respectively.  We count the level $i$ of a tree from the top and in our notations $S_{{\cal K}-1}^{\mathcal L }\equiv s_1=\frac12$. We also introduce the notations $\tilde{S}_i^{\mathcal L}$ and $\tilde{S}_i^{\mathcal R}$ for the corresponding spins in the left-most and right-most branches of the second tree.  By the selection rules  (see Sec. \ref{SubSec:Symmetry}) $\tilde{S}^\eta_i=S_i^\eta\,, S_i^\eta\pm 1$ and the differences $\Delta S^\eta_i\equiv S^\eta_i-\tilde{S}^\eta_i$ can only take values $-1$, $0$ and $1$. 

 We denote the spin at the top of both trees by $S_0$ (this has to be the same in order for there to be a non-trivial matrix element). In most of this supplemental material we set $S_0=0$, implying that 
$S_1^{\mathcal R}=S_1^{\mathcal L}$ and  $\tilde{S}_1^{\mathcal R}=\tilde{S}_1^{\mathcal L}$. This assumption is relaxed in Sec.~\ref{Sec:S0}.
Finally, we denote by $J^{\mathcal L}_i$ and  $J^{\mathcal R}_i$, $i=2\ldots {\cal K}-1$, the spins in the branches next to that of left-most and right-most branches 
. They are shared by both states under consideration in order for the matrix element to be potentially non-zero.  Note that the notations of the present Supplemental material are slightly different from that of the main text. Specifically, in the notations of the manuscript, $S^{\mathcal R }_i$ corresponds to $S^{({\cal K}-i)}_1$ while  $S^{\mathcal L }_i$ to  $S^{({\mathcal K}-i)}_{2^i}$.

\mysubsection{Structure and summary of supplemental material}

This supplemental material is organzied as follows: in Sec.~\ref{Sec:TwoStates} we derive recursion relations that allow us to compute the matrix element for any given pair of states; in Sec.~\ref{Sec:StatisticsOverBaseState}  we convert those into recursion relations for the distribution function of the matrix elements connecting a given base state to other states allowed by the selection rules; in Sec. \ref{Sec:Statistics} we analyze the statistics of the random quantities $\overline{\zeta}$ and $\sigma$ in the large-$L$ limit within the sector of the theory with $S_0=0$ and show that the typical matrix element possesses the scaling
\begin{equation}
V_{\rm typ.}\propto L^{-\beta}\,, \qquad \beta=\frac{17+4\ln 2}{9\ln 2}.
\end{equation}
Finally, in Sec. \ref{Sec:S0} we demonstrate that our results for the asymptotic scaling of the matrix elements hold for arbitrary values of $S_0$.

\mysection{Matrix element for a given pair of states}
 \label{Sec:TwoStates}
\mysubsection{Preliminary symmetry considerations}
\label{SubSec:Symmetry}
Let us first consider a system of spins $(1,2,\cdots,L-1,L)$ interacting via rotationally invariant interactions, that is, we consider a system with $SU(2)$ symmetry, and derive certain constraints on matrix elements of eigenstates due solely to symmetry considerations.  This scenario encompasses the tree-like states that we have considered, but is more general than that.  

Due to the $SU(2)$ symmetry, the eigenstates of the system form multiplets
\begin{equation}
 |S, M\rangle,\qquad M=-S,\ldots S  
   \end{equation}
labeled by the spin $S$.  Next we consider two such multiplets and construct a matrix
\begin{equation}
g^\alpha_{\tilde{M}, M}(\tilde{S}, S)=\langle \tilde{S},\tilde{M}| s_1^\alpha |S, M\rangle,
\end{equation}
where $s_1^\alpha$ is the $\alpha$-spin operator of the first site.

Importantly,  the form of $g^\alpha_{\tilde{M}, M}$ is fully fixed by symmetry.
Let us show first that  $g^\alpha_{\tilde{M}, M}(\tilde{S}, S)$  vanishes whenever $|S-\tilde{S}|>1$ or $S=\tilde{S}=0$. 
Indeed, the matrix element $g^\alpha_{\tilde{M}, M}$  can be considered as a linear function on the triple $(~|\tilde{S}, \tilde{M}\rangle\,, |S, M\rangle\,, s_1^\alpha ) $ which transforms under the $\tilde{S} \otimes S \otimes(s=1)$ representation of the $SU(2)$ group. However since the matrix element $g^\alpha_{\tilde{M}, M}$  is invariant under simultaneous rotations of all the three components of the triplet, i.e.~it transforms trivially under the $SU(2)$ group, the only way for this matrix element to be potentially non-trivial is such that the tensor product of the different representations $\tilde{S} \otimes S \otimes (s=1)$ contains the trivial representation $0$ as a direct summand. Exploiting the rules for angular momentum addition, we see that this can happen only for the cases $|S-\tilde{S}|\leq1$ and $\min(S, \tilde{S})>0$. 

We now proceed to determine the form of the matrix $g^\alpha_{\tilde{M}, M}$.
 For the case $S=\tilde{S}$ obvious symmetry arguments imply that
\begin{equation}
g^\alpha_{\tilde{M}, M}(S, S)\propto S^\alpha_{\tilde{M}, M}
\end{equation} 
where $S^\alpha_{\tilde{M}, M}$ are the spin operators in the spin-S representation. 
Introducing $g^\pm=g^x\pm i g^y$ we write explicitly
\begin{equation}
g^+_{\tilde{M}, M}=\frac{\xi}{\sqrt{\gamma(S, 0)}} \delta_{\tilde{M}, M+1}S^+_{M}\,, \qquad g^+_{\tilde{M}, M}=\frac{\xi}{\sqrt{\gamma(S, 0)}} \delta_{\tilde{M}, M-1}S^-_{M}, \qquad g^z_{\tilde{M}, M}=\frac{\xi}{\sqrt{\gamma(S, 0)}} \delta_{\tilde{M}, M}M.
\label{Eq:gss}
\end{equation}
Here 
\begin{equation}
 S^-_{M}=\sqrt{(S+M)(S-M+1)}\,, \quad   S^+_{M}=\sqrt{(S-M)(S+M+1)},
\end{equation}
and we have also introduced a convenient  normalization factor
\begin{equation}
\gamma(S, S-\tilde{S}=0)=2 S (S+1).
\end{equation}
$\xi$ is a numerical coefficient that depends on the precise internal structure of the states in the multiplets and is where the physics lies (i.e.~details not able to be determined by only symmetry considerations). Different structures give different numerical values; our aim in the supplemental material, which is to understand the scaling of the typical matrix element $V$, thus amounts to understanding $\xi$ for the tree-like states we are considering.

The case of  $|S-\tilde{S}|=1$ is slightly less obvious. However one  can employ   the  Wigner-Eckart theorem \cite{Wigner-Eckart} to find (we assume $S=\tilde{S}+1$; the case   $S=\tilde{S}-1$ is related to the present one by the Hermitian conjugation and the interchange of $S$ and $\tilde{S}$)
\begin{equation}
g^+_{\tilde{M}, M}=\frac{\xi}{\sqrt{\gamma(S, 1)}}\delta_{\tilde{M}, M+1}u^+_{M}\,, \qquad g^-_{\tilde{M}, M}=\frac{\xi}{\sqrt{\gamma(S, 1)}}\delta_{\tilde{M}, M-1}u^-_{M}, \qquad g^z_{\tilde{M}, M}=\frac{\xi}{\sqrt{\gamma(S, 1)}}\delta_{\tilde{M}, M}u^z_{M}.
\label{Eq:gsps}
\end{equation}
Here 
\begin{equation}
u^z_{M}=\sqrt{S^2-M^2}, \qquad u^+_{M}=\sqrt{(S-M)(S-M-1)}\,, \qquad u^-_{M}=
-\sqrt{(S+M)(S+M-1)},
\end{equation}
and 
\begin{equation}
\gamma(S, S-\tilde{S}=1)=2 S \sqrt{4 S^2-1}.
\end{equation} 

\mysubsection{Recursion relations}
\mysubsubsection{Top level}
The symmetry considerations of the previous section allow one to establish recursion relations for the determination of the matrix element. 
Let  us consider first the top level of our two trees. We have for the matrix element
\begin{multline}
V=\langle S_0, S_0| s^\alpha_1 s^\alpha_L| S_0, S_0 \rangle\\=
\sum_{\substack{m+M=S_0\\\tilde{m}+\tilde{M}=S_0}} C^*(\tilde{S}_1^{\mathcal  L}, \tilde{m}, \tilde{S}_1^{\mathcal  R}, \tilde{M}, S_0, S_0) C(S_1^{\mathcal L}, m, S_1^{\mathcal R}, M, S_0, S_0)\langle \tilde{S}_1^{\mathcal L}, \tilde{m} |s_1^\alpha
|S^{\mathcal L}_1, m\rangle  \langle \tilde{S}_1^{\mathcal R}, \tilde{M} |s_L^\alpha
|S^{\mathcal R}_1, M\rangle.
\label{Eq:Vbasic}
\end{multline}
Here $C(s_1, m_1, s_2, m_2, S, M)$ are the Clebsch-Gordan coefficients.  We now use the decomposition of the matrix elements given in the previous section and write  
\begin{eqnarray} 
\langle \tilde{S}_1^{\mathcal L}, \tilde{m} |s_1^\alpha
|S^{\mathcal L}_1, m\rangle=\xi_1^{\mathcal L}\overline{g}^\alpha_{\tilde{m}, m}(\tilde{S}_1^{\mathcal L}, S_1^{\mathcal L} ),\\
\langle \tilde{S}_1^{\mathcal R}, \tilde{m} |s_1^\alpha
|S^{\mathcal R}_1, m\rangle=\xi_1^{\mathcal R}\overline{g}^\alpha_{\tilde{m}, m}(\tilde{S}_1^{\mathcal R}, S_1^{\mathcal R} ),
\end{eqnarray}
to find
\begin{equation}
V=\xi_1^{\mathcal L}\xi_1^{\mathcal R}
\sum_{\substack{m+M=S_0\\\tilde{m}+\tilde{M}=S_0}} C^*(\tilde{S}_1^{\mathcal  L}, \tilde{m}, \tilde{S}_1^{\mathcal  R}, \tilde{M}, S_0, S_0) C(S_1^{\mathcal L}, m, S_1^{\mathcal R}, M, S_0, S_0)\overline{g}^\alpha_{\tilde{m}, m}(\tilde{S}_1^{\mathcal L}, S_1^{\mathcal L} )\overline{g}^\alpha_{\tilde{m}, m}(\tilde{S}_1^{\mathcal R}, S_1^{\mathcal R} ),
\label{Eq:VBasic1}
\end{equation}
where we have defined $\overline{g}$ to be a `normalized' g, i.e.~$\overline{g}$ is given by Eqs. (\ref{Eq:gss}) and (\ref{Eq:gsps}) with $\xi=1$.

Using the explicit expressions  for $\overline{g}$ 
together with the explicit expressions for the  Clebsch-Gordan coefficients, we can perform the summation in Eq.~(\ref{Eq:VBasic1}) and reduce the evaluation of the matrix element to the evaluation of the two factors $ \xi_1^{\eta}$, $\eta = \mathcal{L}, \mathcal{R}$. After long, tedious algebra, we find
\begin{equation}
V=\frac{\xi_1^{\mathcal L}\xi_1^{\mathcal R}}{\sqrt{\gamma(S^{\mathcal L}, \Delta S^{\mathcal L}) \gamma(S^{\mathcal R}, \Delta S^{\mathcal R})}} R_{\Delta S^{\mathcal L}_1, \Delta S^{\mathcal R}_1}(S_1^{\mathcal L}, S_1^{\mathcal R}, S_0).
\label{Eq:VGen}
\end{equation}
We explain the notation of the previous line. For a tree with some $(S_1^{\mathcal L}, S_1^{\mathcal R}, S_0)$, there are $3 \times 3$ possible trees it can connect to, which are given by $\Delta S^\eta_1=S^\eta_1-\tilde{S}^{\eta}_1 = -1, 0, +1$, for $\eta = \mathcal{L}, \mathcal{R}$. Thus, we can compactly store this information in a $3 \times 3$ matrix $R_{\Delta S^{\mathcal L}_1, \Delta S^{\mathcal R}_1}$:
\begin{multline}
 R_{\Delta S^{\mathcal L}_1, \Delta S^{\mathcal R}_1}(S_1^{\mathcal L}, S_1^{\mathcal R}, S_0)  =\\
\frac12\left(
\begin{array}{ccc}
-\sqrt{(j_3+1)(j_3+2)(\Sigma+2)(\Sigma+3)}& -\sqrt{(j_1+1)j_2(j_3+1)(\Sigma+2)} & 
\sqrt{(j_1+1)(j_1+2)(j_2-1)j_2}\\
\sqrt{j_1(j_2+1)(j_3+1)(\Sigma+2)} & \left[j_1 j_2 -(2+j_1+j_2)j_3 -j_3^2\right]/2 & \sqrt{(j_1+1)j_2 j_3(\Sigma+1)}\\
\sqrt{(j_1-1)j_1(j_2+1)(j_2+2)} &-\sqrt{j_1(j_2+1)j_3(\Sigma+1)}&-\sqrt{(j_3-1)j_3\Sigma(\Sigma+1)}
\end{array}
\right)
\end{multline}
with 
\begin{equation}
j_1=S_0+S^{\mathcal L}_1-S^{\mathcal R}_1, \qquad j_2=S_0-S^{\mathcal L}_1+S^{\mathcal R}_1\,, \qquad  j_3=-S_0+S^{\mathcal L}_1+S^{\mathcal R}_1\,, \qquad \Sigma=j_1+j_2+j_3.
\end{equation}

Equation (\ref{Eq:VGen}) takes an especially simple form for the case $S_0 = 0 $ where $S^{\mathcal L}= S^{\mathcal R}$ and   $\Delta S^{\mathcal L}= \Delta S^{\mathcal R}$. In this limit (which we are going to focus on in the following) we find
\begin{equation}
V=-\frac{1}{2}\xi_1^{\mathcal L}\xi_1^{\mathcal R}.
\label{Eq:VS0}
\end{equation}

\mysubsubsection{The factor $\xi$}
In the previous section, we reduced the problem to the evaluation of the two factors $\xi_1^{\mathcal L}$ and $\xi_1^{\mathcal R}$ coming from the left-most and right-most branches of a tree. They can be computed independently and in this section we focus on $\xi_1^{\mathcal L}$ omitting the superscript ${\mathcal L}$. 

To establish the recursion relations let us consider the first level of the tree (just below the top node). We have
\begin{multline}
\xi_1 \overline{g}^{\alpha}_{\tilde{M}, M}(\tilde{S}_1,S_1)\equiv\langle \tilde{S}_1,\tilde{M}| s_1^\alpha | S, M\rangle=
\sum_{\substack{
m+n=M \\
\tilde{m}+\tilde{n}=\tilde{M}}} C^*(\tilde{S}_2, \tilde{m}, J_2, \tilde{n}, \tilde{S}_1, \tilde{M})C(S_2, m, J_1, n, S_1, M)
\langle \tilde{S}_2, \tilde{m}|s_1^\alpha|S_2, m\rangle \delta_{n, \tilde{n}}\\=
\xi_2
\sum_{\substack{
m+n=M \\
\tilde{m}+\tilde{n}=\tilde{M}}} C^*(\tilde{S}_2, \tilde{m}, J_2, \tilde{n}, \tilde{S}_1, \tilde{M})C(S_2, m, J_1, n, S_1, M)
\overline{g}^\alpha_{\tilde{m}, m}(\tilde{S}_2, S_2)\delta_{n, \tilde{n}}.
\end{multline} 

Now, knowledge of $\overline{g}$ allows us to perform the summation explicitly and find the recursion relation for the factor $\xi$ (while we are currently working on the first level of the tree; it is obvious that the same procedure can be pushed further down the tree)
\begin{equation}
\xi_{k-1}=\frac{\xi_k}{\sqrt{\gamma(S_{k-1}, \Delta_{S_{k-1}}) \gamma(S_{k}, \Delta {S_{k}}})}f_{\Delta S_{k-1}, \Delta S_k}(S_{k-1}, S_k, J_k).
\label{Eq:xiExact}
\end{equation}
Here
\begin{equation}
\gamma(S, \Delta S)=\left\{ \begin{array}{cc}
2(S+1)\sqrt{(2S+1)(2S+3)}\,,  \quad& \Delta S=-1\\
2S(S+1)\,, \qquad& \Delta S=0 \\
2S\sqrt{4S^2-1}\,, \quad& \Delta S=1\\
\end{array}\right. 
\label{Eq:gammaFinal}
\end{equation}
and we have stored the information in a $3 \times 3$ matrix $f_{\Delta S_1, \Delta S_2}$:
\begin{multline}
f_{\Delta S_1, \Delta S_2}(S_1, S_2, J_2) =\\
\left(
\begin{array}{ccc}
\sqrt{(j_1+2)(j_1+1) (\Sigma+2)(\Sigma+3)}& \sqrt{(j_1+1) (j_2+1) j_3(\Sigma+2)} & 
-\sqrt{(j_2+1)(j_2+2)(j_3-1)j_3}\\
-\sqrt{(j_1+1)j_2(j_3+1)(\Sigma+2)} &  \left[j_1(\Sigma+2)-j_2  j_3\right]/2 &  -\sqrt{j_1(j_2+1)j_3(\Sigma+1)}\\
-\sqrt{j_2(j_2-1)(j_3+1)(j_3+2)} & \sqrt{j_1 j_2 (1+j_3)(1+\Sigma )}& \sqrt{j_1(j_1-1)\Sigma (\Sigma+1)}
\end{array}
\right)
\label{Eq:fExact}
\end{multline}
with
\begin{equation}
j_1=S_1+S_2-J_2\,, \qquad j_2=S_1-S_2+J_2\,, \qquad j_3=-S_1+S_2+J_2\,, \qquad 
\Sigma=j_1+j_2+j_3.
\label{Eq:j}
\end{equation}

Equations (\ref{Eq:xiExact}) and (\ref{Eq:VS0}) together with Eqs. (\ref{Eq:gammaFinal}), (\ref{Eq:fExact}) and (\ref{Eq:j}) constitute the main result of this section. 
They can be used for efficient numerical evaluation of the matrix elements.

In the limit of  large system size and for the situation when the level considered is not too close to the bottom of the tree, all spins involved are large and differ significantly one from another, one can significantly simplify the recursion relations. One finds
\begin{equation}
\xi_{k-1}=\tilde{f}_{\Delta S_{k-1}, \Delta S_k}(S_{k-1}, S_k, J_k)\xi_{k},
\label{Eq:xiLarge}
\end{equation}
where
\begin{equation}
\tilde{f}_{\Delta S_1, \Delta S_2}(S_1, S_2, J_2)=
\frac{1}{4S_{1} S_{2}}\left(
\begin{array}{ccc}
U& \sqrt{2UV} & 
-V\\
-\sqrt{2UV} &  U-V &  -\sqrt{2UV}\\
-V & \sqrt{2UV}& U
\end{array}
\right)\,, \qquad U=j_1\Sigma,\qquad V=j_2 j_3.
\label{Eq:fLarge}
\end{equation}

We will employ Eqs. (\ref{Eq:xiLarge}) and (\ref{Eq:fLarge}) in our asymptotic analysis of the matrix elements for large systems. 

Before closing this section let us draw the attention of the reader to the fact that according to Eq. (\ref{Eq:fExact}), matrix elements can still vanish despite being allowed by the selection rules . It is easy to see that the factor $\xi$ vanishes for the transitions with $\Delta S_{k-1}=\Delta S_k=0$ provided that
the three spins $S_{{\cal K}-1}$, $S_k$ and $J_k$ form a right-angled triangle
\begin{equation}
 S_{k-1}(S_{k-1}+1)+ S_{k}(S_{k}+1)-J_{k}(J_{k}+1)=0.
 \label{Eq:triangle}
 \end{equation}
However, these vanishing matrix elements do not influence the scaling of the connectivity with the system size presented in the main text, because the density of the configurations satisfying 
Eq. (\ref{Eq:triangle}) vanishes in the limit of large system size.

\mysection{Statistics over a fixed base state}. 
\label{Sec:StatisticsOverBaseState}
Let us now suppose that we choose randomly a a tree-like state (with top spin zero) and consider the distribution of non-zero matrix elements connecting this states to its neighbors.
More precisely we are interested in the distribution function $\zeta=\log_2 |V|$
In this section we show how this distribution can be computed. 

\mysubsection{Top level}
We describe the states connected to the fixed base state by the set of $\Delta S^{\eta}_k=S^{\eta}_k-\tilde{S}_k^\eta$. 
Let us introduce the generating function for the random variable $\zeta$
\begin{equation}
F_\zeta(\lambda)=\langle e^{i\lambda \zeta}\rangle=\frac{1}{K}\sum_{\rm all\; connected\; states}e^{i\lambda \zeta}=\frac{e^{-i\lambda}}{\cal K}
\sum_{\rm all\; connected\; states} e^{i\lambda (\omega^{\cal R}_1 + \omega^{\cal L}_1)}.
\label{Eq:Fdef}
\end{equation}
Here  $K$ is the number of states connected to the fixed base state, $\omega^\eta_1=\log_2 |\xi^\eta_1|$ and we have used Eq. (\ref{Eq:VS0}).
The summation goes over all the states for which the base state has non-zero matrix element with. This condition is local in $\Delta S$ in the sense that it can be expressed by an indicator function
\begin{equation}
\delta_{S_1^{\mathcal R} S_1^{\mathcal L}}\prod_{\eta={\mathcal R}, {\mathcal L}}\prod_{k={1}}^{\mathcal K-2}\psi_{J^\eta_{k+1}, S^\eta_{k+1}, S^\eta_k}(\Delta S^\eta_{k+1}, \Delta S^\eta_k)
\end{equation}
where the  Kronecker symbol is due to the fact that  we are considering trees with zero spin on top and the function 
 $\psi_{J_2, S_2, S_1}(\Delta S_1, \Delta S_2)$ is only non-zero if the following conditions are fulfilled simultaneously
 \begin{enumerate}
 \item $S_2-\Delta S_2\geq 0$;
 \item $|J_2-S_2+\Delta S_2|\leq S_1-\Delta S_1\leq J_2+S_2-\Delta S_2 $;
 \item $ S_{1}(S_{1}+1)+ S_{2}(S_{2}+1)-J_{2}(J_{2}+1)\neq 0 $ if $\Delta S_1 =\Delta S_2=0$. 
 \end{enumerate}

Note further that $\omega^{\eta}_1$ is a function of the corresponding $\Delta S^{\eta}$ only. Thus we can rewrite Eq. 
(\ref{Eq:Fdef})  as
\begin{equation}
F_\zeta(\lambda)=\frac{e^{i\lambda}}{K}\sum_{\Delta S}F_{\omega_1}^{\mathcal R}(\lambda, \Delta S) F_{\omega_1}^{\mathcal L}(\lambda, \Delta S)
\label{Eq:FzetaFund}
\end{equation}
 with 
 \begin{equation}
 F_{\omega_1}^{\eta}(\lambda, \Delta S)=\sum_{\Delta S^\eta_{k=1, \ldots {\cal K}-2}}
 \delta_{\Delta S^\eta_1, \Delta S}
 \exp\left[i\omega^\eta_1\right] \prod_{k={1}}^{\mathcal K-2}\psi_{J^\eta_{k+1}, S^\eta_{k+1}, S^\eta_k}(\Delta S^\eta_{k+1}, \Delta S^\eta_k).
 \end{equation}
 
\mysubsection{Recursion relations}
 The functions $F_\omega^{{\mathcal R}({\mathcal L})}$ introduced in the previous section  satisfy recursion 
 relations that we are going to derive now. We omit the irrelevant superscript $\eta$ and rephrase the recursion relation (\ref{Eq:xiExact}) in terms of $\omega=\log_2|\xi|$
 \begin{equation}
 \omega_{1}=\omega_2+ r_{S_1, S_2, J_2}(\Delta S_1, \Delta S_2)\,, \qquad r_{S_1, S_2, J_2}(\Delta S_1, \Delta S_2)=\log_2\frac{f_{\Delta S_1, \Delta S_2}(S_1, S_2, J_2)}{\sqrt{\gamma(S_1, \Delta S_1) \gamma(S_2, \Delta S_2)}}.
 \label{Eq:omegaExact}
 \end{equation}
 We immediately get
 \begin{equation}
 F_{\omega_1}(\Delta S_1, \lambda)=\sum_{\Delta S_2} F_{\omega_2}(\Delta S_2, \lambda)\psi_{J_2, S_2, S_1}(\Delta S_1,\Delta S_2)\exp\left[i\lambda r_{S_1, S_2, J_2}(\Delta S_1, \Delta S_2)\right].
 \label{Eq:Fomega}
 \end{equation}

 \mysubsection{Mean and dispersion}
 The random process (\ref{Eq:omegaExact}) can be regarded as a kind of random walk with short-range correlations. Therefore, we can expect the probability distribution for the variable $\omega$ to approach a Gaussian form in the large-$L$ limit.  
 This expectation is also supported by numerical simulations described in the main text. Thus, instead of analyzing the full generating functions  $F^{\eta}_\omega$ and $F_\zeta$ [see Eq. (\ref{Eq:FzetaFund})] we can limit ourself to the consideration of the mean value $\overline{\zeta}$ and dispersion $\sigma$ of the variable $\zeta$. 
 
To that end, we introduce the power series expansion of the functions $F_{\omega_1}^{\eta}(\Delta S, \lambda)$
 \begin{equation}
 F_{\omega_1}^\eta(\Delta S, \lambda)=F_{\omega_1}^{\eta, (0)}(\Delta S)+i\lambda F_{\omega_1}^{\eta, (1)}(\Delta S)-\frac{\lambda^2}{2}F_{\omega_1}^{\eta, (2)}(\Delta S)+\ldots.
 \end{equation}
In terms of $F_{\omega_1}^{\eta, (i)}(\Delta S)$ we get
\begin{eqnarray}
K&=&\sum_{\Delta S}F_{\omega_1}^{{\mathcal L}, (0)}(\Delta S) F_{\omega_1}^{{\mathcal R}, (0)}(\Delta S),
\label{Eq:KExact}
\\
\overline{\zeta}&=&-1 +\frac1K\sum_{\Delta S}\left[ F_{\omega_1}^{{\mathcal R}, (0)}(\Delta S) F_{\omega_1}^{{\mathcal L}, (1)}(\Delta S) 
+ F_{\omega_1}^{{\mathcal R}, (1)}(\Delta S) F_{\omega_1}^{{\mathcal L}, (0)}(\Delta S)\right],
\label{Eq:xibarExact}
\\
\sigma^2&=&\frac1K \sum_{\Delta S}\left[ 
F_{\omega_1}^{{\mathcal R}, (0)}(\Delta S) F_{\omega_1}^{{\mathcal L}, (2)}(\Delta S)+
F_{\omega_1}^{{\mathcal R}, (2)}(\Delta S) F_{\omega_1}^{{\mathcal L}, (0)}(\Delta S)+
2F_{\omega_1}^{{\mathcal R}, (1)}(\Delta S) F_{\omega_1}^{{\mathcal L}, (1)}(\Delta S)   
\right]-(\overline{\zeta} +1)^2.\quad
\label{Eq:sigmaExact}
\end{eqnarray}

The recursion relations for $F_{\omega_1}^{\eta, (i)}(\Delta S)$  can be read off from Eq. (\ref{Eq:Fomega})
\begin{eqnarray}
F_{\omega_1}^{(0)}( \Delta S_1)&=&
\sum_{\Delta S_2} 
\psi_{J_{2}, S_{2}, S_{1}}(\Delta S_{1}, \Delta S_2)F^{(0)}_{\omega_2}( \Delta S_2),
\label{Eq:F0Exact}
\\
F_{\omega_1}^{(1)}( \Delta S_1)&=&
\sum_{\Delta S_2} 
\overline{\psi}_{J_{2}, S_{2}, S_{1}}(\Delta S_{1}, \Delta S_2)\left[F^{(1)}_{\omega_2}( \Delta S_2)+F^{(0)}_{\omega_2}( \Delta S_2)r(\Delta S_1, \Delta S_2)\right],
\label{Eq:F1Exact}
\\
F_{\omega_1}^{(2)}( \Delta S_1)&=&
\sum_{\Delta S_2} 
\overline{\psi}_{J_{2}, S_{2}, S_{1}}(\Delta S_{1}, \Delta S_2)\left[F^{(2)}_{\omega_2}( \Delta S_2)+2F^{(1)}_{\omega_2}( \Delta S_2)r(\Delta S_1, \Delta S_2)+F^{(0)}_{\omega_2}( \Delta S_2)r^2(\Delta S_1, \Delta S_2)\right].\qquad\quad
\label{Eq:F2Exact}
\end{eqnarray}
Here we suppress the superscript $\eta$ and use the shorthand notation $r(\Delta S_1, \Delta S_2)\equiv r_{S_1, S_2, J_2}(\Delta S_1, \Delta S_2)$. 

Equations (\ref{Eq:KExact}), (\ref{Eq:xibarExact}) and (\ref{Eq:sigmaExact}) together with  (\ref{Eq:F0Exact}), (\ref{Eq:F1Exact}) and (\ref{Eq:F2Exact}) constitute the main result of this section. They are exact and we have tested them against direct numerical simulations. In the next section we will specialize to the case of large system sizes.

\mysubsection{Large-$L$ limit}
In the large-$L$ limit constraints imposed by the indicator functions $\psi(\Delta S_1, \Delta S_2)$ are irrelevant.  Equation (\ref{Eq:F0Exact}) implies then that 
$F^{(0)}_{\omega_1}$ is independent of $\Delta S$ and satisfies
\begin{equation}
F_{\omega_1}^{(0)}=3 F_{\omega_2}^{(0)}
\end{equation}
which with together with Eq. (\ref{Eq:KExact}) leads to the scaling of the connectivity
\begin{equation}
K\propto 3^{2 {\cal K}}=L^{\alpha}, \qquad \alpha=2\log_2 3.
\end{equation}

It is convenient to introduce $\overline{F}^{(i)}_{\omega_1}(\Delta S)= F^{(i)}_{\omega_1}(\Delta S)/ F^{(0)}_{\omega_1}$, $i=1, 2$. The recursion relations (\ref{Eq:F1Exact}) and (\ref{Eq:F2Exact}) now acquire the form
\begin{eqnarray}
\overline{F}_{\omega_1}^{(1)}( \Delta S_1)&=&
\frac13\sum_{\Delta S_2} 
\left[\overline{F}^{(1)}_{\omega_2}( \Delta S_2)+r(\Delta S_1, \Delta S_2)\right],
\label{Eq:Fbar1}
\\
\overline{F}_{\omega_1}^{(2)}( \Delta S_1)&=&
\frac13\sum_{\Delta S_2} 
\left[\overline{F}^{(2)}_{\omega_2}( \Delta S_2)+2\overline{F}^{(1)}_{\omega_2}( \Delta S_2)r(\Delta S_1, \Delta S_2)+ r^2(\Delta S_1, \Delta S_2)\right].
\label{Eq:Fbar2}
\end{eqnarray}

It proves useful to introduce
\begin{equation}
E_{\omega_k}=\sum_{\Delta S}\overline{F}_{\omega_k}^{(1)}(\Delta S)\,, \qquad 
G_{\omega_k}=\sum_{\Delta S}\overline{F}_{\omega_k}^{(2)}(\Delta S)-\frac13E_{\omega_k}^2
\end{equation}
satisfying recursion relations 
\begin{eqnarray}
E_{\omega_1}&=&E_{\omega_2}+\frac{1}{3}u,
\label{Eq:ERecFinal}
\quad\\
G_{\omega_1}&=&G_{\omega_2}+\sum_{\Delta S_1, \Delta S_2}\left[\overline{F}^{(1)}_{\omega_2}(\Delta S_2)-\frac13 E_{\omega_2}\right]r(\Delta S_1, \Delta S_2)-
\frac{1}{27}u^2+\frac13\sum_{\Delta S_1\Delta S_2}r^2(\Delta S_1, \Delta S_2).\quad
\label{Eq:GRecFinal}
\end{eqnarray}
Here 
\begin{equation}
u=u(S_1, S_2, J_2)=\sum_{\Delta S_1, \Delta S_2}r_{S_1, S_2, J_2}(\Delta S_1, \Delta S_2).
\label{Eq:uDef}
\end{equation}

Expressions (\ref{Eq:xibarExact}) and  (\ref{Eq:sigmaExact}) for the mean and dispersion can now also be simplified. We get after straightforward but lengthy algebra
\begin{eqnarray}
\overline{\zeta}&=&-1 +\frac13\left[  E^{\mathcal L}_{\omega_1}+ E^{\mathcal R}_{\omega_1}\right]\,, 
\label{Eq:xibarFinal}
\\
\sigma^2 &=&
\frac13 G^{\mathcal L}_{\omega_1}+ \frac13 G^{\mathcal R}_{\omega_1}+\frac{2}{27}
E^{\mathcal R}_{\omega_1}u^{\mathcal L}+ \frac{2}{27}
E^{\mathcal L}_{\omega_1}u^{\mathcal L}+\frac{2}{27}\sum_{\Delta S_1}v^{\mathcal L}(\Delta S_1) v^{\mathcal R}(\Delta S_1).
\label{Eq:sigmaFinal}
\end{eqnarray}
Here
\begin{equation}
v(\Delta S_1)=\sum_{\Delta S_2} r(\Delta S_1, \Delta S_2).
\end{equation}

\mysection{Statistics of $\overline{\zeta}$ and $\sigma$}
\label{Sec:Statistics}
In this section we turn to the main subject of this Supplemental Material.  In Sec. \ref{Sec:StatisticsOverBaseState} we have characterized the distribution of  the matrix elements connecting fixed tree-like state to other states by the $\overline{\zeta}$ and $\sigma$, the mean and the standard deviation of the logarith of the matrix element. Given different base states the quantities $\overline{\zeta}$ and $\sigma$  are themselves  random variables. We would like to characterize their statistics by computing the averages $\langle \xi\rangle$ and $\langle \sigma\rangle$ with respect to the chosen base state as well as the ``mesoscopic fluctuations'' of $\overline{\zeta}$, $\langle (\delta \overline{\zeta})^2\rangle$.

\mysubsection{Averaging of $\overline{\zeta}$}
\label{SubSec:xiAv}
\mysubsubsection{Recursion relations in large-$L$ limit.}
We start with the computation of $\langle \overline{\zeta}\rangle$. We limit ourself to the analysis in the large-$L$ limit where, according to Eq. (\ref{Eq:xibarFinal}),
\begin{equation}
\langle \overline{\zeta}\rangle\approx \frac{2}{3}\langle E^{\mathcal L}_{\omega_1}\rangle.
\label{Eq:xiaE}
\end{equation}
The quantity $E^{\mathcal L}_{\omega_1}$ satisfies the recursion relation (as usual we omit the superscript and restore the dependence on the spins in the base state, suppressed previously)
\begin{equation}
E_{\omega_1}=E_{\omega_2}+\frac{1}{3}u(S_1, S_2, J_2)
\label{Eq:ERecFinalExp}
\end{equation}
with $u(S_1, S_2, J_2)$ given by [see Eqs. (\ref{Eq:uDef}), (\ref{Eq:omegaExact}) and (\ref{Eq:fLarge})]
\begin{equation}
u(S_1, S_2, J_2)=\log_2 \frac{U^4 V^4 |U-V|}{2^{16} S_1^9 S_2^9}\,, \qquad U=(S_1+S_2)^2-J_2^2\,,\qquad V=J_2^2-(S_1-S_2)^2. 
\label{Eq:uLarge}
\end{equation}

Let us introduce the number of trees of depth $k$ with spin $S$ on top, $N_k(S)$. Simple combinatorics gives
\begin{equation}
N_k(S)=\frac{L_k ! (2S+1)}{\left(\frac{L_k}{2}-S\right)! \left(\frac{L_k}{2}+S+1\right)!}.
\end{equation}
Here $L_k=2^{k-1}$ is the number of spins underlying a tree of depth $k$. 

For the average we are interested in we can now write
\begin{equation}
\langle E_{\omega_1}\rangle=\frac{1}{N_{\cal K}(0)}\sum_{\{S, J\}} E_{\omega_1}(\{S, J\}) N_{{\cal K}-1}(S_1)\prod_{k=2}^{{\cal K}-2} N_{{\cal K}-k}(J_k).
\label{Eq:EaBasic}
\end{equation} 
Here the summation goes over all consistent sets of $S_{k=1,\ldots  {\cal K}-1}$ and $J_{k=2, \ldots {\cal K}-1}$. 

 We introduce now
\begin{equation}
R_{\omega_k}(S)=\frac{1}{N_{\cal K}(0)}\sum_{\{S, J\}_k} E_{\omega_k}(\{S, J\}) \delta_{S_k, S}\prod_{k=2}^{{\cal K}-k-1} N_{{\cal K}-k}(J_k)
\label{Eq:Rdef}
\end{equation}
and rewrite (\ref{Eq:EaBasic}) as
\begin{equation}
\langle E_{\omega_1}\rangle=\frac{1}{N_{\cal K}(0)}\sum_{S_1=0}^{L_{\cal K}/2} N_{{\cal K}-1}(S_1)R_{\omega_1}(S_1).
\label{Eq:EaBasic1}
\end{equation} 
In Eq. (\ref{Eq:Rdef}) the summation runs over all consistent sets of $S_{i=k, \ldots {\cal K}-1}$ and $J_{i=k+1, \ldots {\cal K}-1}$. 

The function  $R_{\omega_1}(S)$ satisfies the recursion relations that can be read off Eq. (\ref{Eq:ERecFinalExp})
\begin{equation}
R_{\omega_{k-1}}(S_{k-1})= 
\sum_{S_k}D_{k}(S_{k-1}, S_k)R_{\omega_k}(S_k)+\frac{1}{3}\sum_{\substack{J_k, S_k\\ |S_k-J_k|\leq S_{k-1}\leq S_k+J_k}}u(S_{k-1}, S_k, J_k)N_{{\cal K}-k}(J_k) N_{{\cal K}-k}(S_k).
\label{Eq:Rrec}
\end{equation}
 with 
 \begin{equation}
 D_{k}(S_{k-1}, S_k)=\sum_{\substack{J_k=|S_k-S_{k-1}|\\J_k\leq L_{{\cal K}-k}/2}}^{S_k+S_{k-1}} N_{{\cal K}-k}(J_k)
 \end{equation}
 
We are now in position to formulate the final set of equations describing $\langle E_{\omega_1}\rangle$ in the large-$L$ limit. 
To achieve this goal we switch to a continuum description of the  spins.  In the continuum limit the function $N_k(S_k)$  is given by
\begin{equation} 
N_k(S)=\sqrt{\frac{2}{\pi}}\frac{2^{L_k+2} S}{L_k^{3/2}}e^{-2S^2/L_k}=\sqrt{\frac{2}{\pi}}\frac{2^{L_{k}}}{L_k}n(S/\sqrt{L_k})\,, \qquad n(s)= 4 s
e^{-2s^2}\label{Eq:ns}
\end{equation} 
We thus introduce the continuum variables $s_k=S_k/\sqrt{L_{{\cal K}-k}}$ and $j_k=J_k/\sqrt{L_{{\cal K}-k}}$  together with functions 
\begin{equation}
\tilde{R}_{\omega_k}(s_k)=\frac{\sqrt{2}L_{{\cal K}-k}}{2^{L_{{\cal K}-k}}}R_{\omega_k}(s\sqrt{L_{{\cal K}-k}})
\end{equation}
and switch from summation to integration in Eq. (\ref{Eq:Rrec}). The result reads
\begin{equation}
\tilde{R}_{\omega_{k-1}}(s_{k-1})=
\int_0^\infty ds_k \tilde{D}(s_{k-1}, s_k) \tilde{R}_{\omega_k}(s_k)\\+
\frac{4\sqrt{2}}{3\pi} \int_{ |j_k-s_k|\leq \sqrt{2} s_{k-1}\leq j_k+s_k}^\infty dj_k d s_k
\tilde{u}(s_{k-1}, s_k, j_k )n(j_k) n(s_k).
\label{Eq:RecFinal}
\end{equation}
Here (see  Eq.~({\ref{Eq:uLarge}}))
\begin{eqnarray}
\tilde{D}(s_{k-1}, s_k)&=&\frac{2\sqrt{2}}{\sqrt{\pi}}\left[e^{-2(s_k+\sqrt{2}s_{k-1})^2}- e^{-2(s_k-\sqrt{2}s_{k-1})^2}\right],\\
\tilde{u}(s_{k-1}, s_k, j_k )&=&u(s_{k-1}\sqrt{L_{{\cal K}-k+1}}, s_k\sqrt{L_{{\cal K}-k}}, j_k \sqrt{L_{{\cal K}-k}})\nonumber\\&=& \log_2 \frac{\left[(\sqrt{2}s_{k-1}+s_k)^2-j_{k}^2\right]^4 \left[j_{k}^2-(\sqrt{2}s_{k-1}-s_k)^2\right]^4 \left|2s_{k-1}^2+s_k^2-j_k^2\right|}{2^{19}\sqrt{2} s_{k-1}^9 s_k^9}.
\end{eqnarray}
Finally, in the continuum limit  Eq. (\ref{Eq:EaBasic1}) takes the form
\begin{equation}
\langle E_{\omega_1}\rangle= \int_0^\infty ds n(s)\tilde{R}_{\omega_1}(s).
\label{Eq:EaFinal}
\end{equation} 

Equations (\ref{Eq:EaFinal}) and  (\ref{Eq:RecFinal}) constitute  the main result of this section. We use them below to derive the asymptotic behavior of the average $\langle\overline{\zeta}\rangle$.

\mysubsubsection{Asymptotic behavior of  $\langle\overline{\zeta}\rangle$}
To find the asymptotic behavior  of $\tilde{R}_{\omega_1}(S)$ we need here several properties of the kernel $\tilde{D}$ in Eq. (\ref{Eq:RecFinal}). It is easy to check that 
the function $n(S)$, Eq.~(\ref{Eq:ns}), is an eigenfunction of $\tilde{D}$ with eigenvalue $1$. Moreover, one can show that all other eigenvalues are smaller than 1 and the corresponding eigenfunctions $\psi_{2n+1}$ can be expressed in terms of the Hermite polynomials
\begin{equation}
\psi_{2n+1}(s)\sim H_{2n+1}(\sqrt{2} s)e^{-2s^2},
\end{equation}
which are orthorgonal on $(0, \infty)$ with weight $e^{2 s^2}$. 

Expanding now $\tilde{R}_{\omega_k}$ in the eigenfunctions of $\tilde{D}$ as
\begin{equation}
R_{\omega_k}(s)=c_k n(s)+\ldots 
\end{equation}
we find from Eq. (\ref{Eq:RecFinal})
\begin{equation}
c_{k-1}=c_k+\frac{4\sqrt{2}}{3\pi |n(s)|^2} \int_{ |j_k-s_k|\leq \sqrt{2} s_{k-1}\leq j_k+s_k}^\infty ds_{k-1} dj_k d s_k
\tilde{u}(s_{k-1}, s_k, j_k )n(j_k) n(s_k)n(s_{k-1})e^{2 s_{k-1}^2}
\end{equation}
with 
\begin{equation}
 |n(s)|^2=\int_0^\infty ds e^{2 s^2}n^2(s)=\sqrt{2\pi}.
\end{equation}

Computation of the integral leads to
\begin{equation}
c_{k-1}=c_k -\frac{17+4\ln 2}{3\sqrt{\pi}\ln 2}
\end{equation}
implying 
\begin{equation}
\tilde{R}_{\omega_1}(s)\sim -\frac{(17+4\ln 2){\cal K}}{3\sqrt{\pi}\ln 2}n(s).
\label{Eq:Rass}
\end{equation}

Equation (\ref{Eq:Rass}) together with Eqs. (\ref{Eq:EaFinal}) and (\ref{Eq:xiaE}) gives us now the central  result of this section.
\begin{equation}
\langle\overline{\zeta}\rangle\sim -\beta \log_2 L\,, \qquad \beta=\frac{17+4\ln 2}{9\ln2}.
\label{Eq:mainAns}
\end{equation} 
This is the scaling stated in the main text.

\mysubsection{Averaging of $\sigma^2$}
\label{SubSec:sigmaAv}
Let us now discuss the averaging of the dispersion $\sigma^2$. Such an averaging can be performed along the lines of Sec. \ref{SubSec:xiAv} on the basis of Eqs. (\ref{Eq:sigmaFinal}), (\ref{Eq:ERecFinal}) and (\ref{Eq:GRecFinal}) that we collect here for the convenience of the reader 
\begin{eqnarray}
\sigma^2 &=&
\frac13 G^{\mathcal L}_{\omega_1}+ \frac13 G^{\mathcal R}_{\omega_1}+\frac{2}{27}
E^{\mathcal R}_{\omega_1}u^{\mathcal L}+ \frac{2}{27}
E^{\mathcal L}_{\omega_1}u^{\mathcal L}+\frac{2}{27}\sum_{\Delta S_1}v^{\mathcal L}(\Delta S_1) v^{\mathcal R}(\Delta S_1),
\label{Eq:sigmaFinalCopy}\\
E_{\omega_1}&=&E_{\omega_2}+\frac{1}{3}u,
\label{Eq:ERecFinalCopy}
\quad\\
G_{\omega_1}&=&G_{\omega_2}+\sum_{\Delta S_1, \Delta S_2}\left[\overline{F}_{\omega_2}(\Delta S_2)-\frac13 E_{\omega_2}\right]r(\Delta S_1, \Delta S_2)-
\frac{1}{27}u^2+\frac13\sum_{\Delta S_1\Delta S_2}r^2(\Delta S_1, \Delta S_2).\quad
\label{Eq:GRecFinalCopy}
\end{eqnarray}

The analysis is quite involved and we do not present it here. Straightforward inspection of Eq. (\ref{Eq:GRecFinalCopy}) shows that $G_{\omega_1}$, very much like $E_{\omega_1}$ discussed in the previous section, grows linearly with the ${\cal K}=\log_2 L$. It is easy to see then from Eq. (\ref{Eq:sigmaFinalCopy}) that
\begin{equation}
\langle\sigma^2\rangle\propto \log_2 L 
\end{equation} 
which is the result stated in the main text.

\mysubsection{Mesoscopic fluctuations of $\overline{\zeta}$}
Let us now study the mesoscopic fluctuations of $\overline{\zeta}$. According to Eq. (\ref{Eq:xibarFinal}) we have
\begin{equation}
\langle(\delta\overline{\zeta})^2\rangle=\frac{2}{9}\left[\left\langle \left(E_{\omega_1}^{\mathcal L}\right)^2\right\rangle +\left\langle E_{\omega_1}^{\mathcal L}E_{\omega_1}^{\mathcal R}\right\rangle -
2\langle E_{\omega_1}^{\mathcal L}\rangle^2\right].
\label{Eq:deltaxiBase}
\end{equation}
We would like to show here that $\langle(\delta\overline{\zeta})^2\rangle$ scales linearly with ${\cal K}=\log_2 L$. 
Since (see Sec. \ref{SubSec:xiAv} )
\begin{equation}
\langle E_{\omega_1}\rangle\sim \frac{3}{2}\beta {\cal K},\qquad \beta=\frac{17+4\ln 2}{9\ln2},
\label{Eq:EaFinalFinal}
\end{equation}
we need to analyze the average
\begin{equation}
\left\langle \left(E_{\omega_1}^{\mathcal L}\right)^2\right\rangle +\left\langle E_{\omega_1}^{\mathcal L}E_{\omega_1}^{\mathcal R}\right\rangle
\end{equation}
to the order ${\cal K}^2$ and show that the quadratic scaling cancels in Eq. (\ref{Eq:deltaxiBase}).

\mysubsubsection{Averaging  $ \left(E_{\omega_1}^{\mathcal L}\right)^2$}
Along the lines of Sec. \ref{SubSec:xiAv} we can write the average in question as
\begin{equation}
\langle E_{\omega_1}^2\rangle=\frac{1}{N_{\cal K}(0)}\sum_{S_1=0}^{L_{\cal K}/2} N_{{\cal K}-1}(S_1)U_{\omega_1}(S_1)
\label{Eq:Ea2Basic1}
\end{equation} 
 with $U_{\omega_k}(S_1)$ given by
 \begin{equation}
U_{\omega_k}(S)=\frac{1}{N_{\cal K}(0)}\sum_{\{S, J\}_k} E^2_{\omega_k}(\{S, J\}) \delta_{S_k, S}\prod_{k=2}^{{\cal K}-k-1} N_{{\cal K}-k}(J_k)
\label{Eq:Udef}
\end{equation} 
 and satisfying recursion relations
 \begin{multline}
U_{\omega_{k-1}}(S_{k-1})= 
\sum_{S_k}D_{k}(S_{k-1}, S_k)U_{\omega_k}(S_k)+
\frac{2}{3}\sum_{\substack{J_k, S_k\\ |S_k-J_k|\leq S_{k-1}\leq S_k+J_k}}u(S_{k-1}, S_k, J_k)N_{{\cal K}-k}(J_k) R_{\omega_k}(S_k)
\\+\frac{1}{9}\sum_{\substack{J_k, S_k\\ |S_k-J_k|\leq S_{k-1}\leq S_k+J_k}}u^2(S_{k-1}, S_k, J_k)N_{{\cal K}-k}(J_k) N_{{\cal K}-k}(S_k).
\label{Eq:UrecBasic}
\end{multline}

As we are interested only in the leading ${\cal K}^2$ term in the asymptotic we can omit term in Eq. (\ref{Eq:UrecBasic}). Going to the continuum description of spins we get
\begin{eqnarray}
\langle E^2_{\omega_1}\rangle&= &\int_0^\infty ds n(s)\tilde{U}_{\omega_1}(s)
\label{Eq:EaSFinal}\\
\tilde{U}_{\omega_{k-1}}(s_{k-1})&=& 
\int_0^\infty d s_k \tilde{D}(s_{k-1}, s_k)\tilde{U}_{\omega_k}(s_k)+
\frac{4}{3}\sqrt{\frac{2}{\pi}}\int_{ |j_k-s_k|\leq \sqrt{2} s_{k-1}\leq j_k+s_k}^\infty dj_k d s_k n(j_k) \tilde{R}_{\omega_k}(s_k)\tilde{u}(s_{k-1}, s_k, j_k).\qquad\qquad
\label{Eq:URecSimple}
\end{eqnarray}

The scaling of $\tilde{R}_{\omega_k}(s_k)$ is give by (cf.~Eq.~(\ref{Eq:Rass}))
\begin{equation}
\tilde{R}_{\omega_k}(s)\sim -\frac{(17+4\ln 2)({\cal K}-k)}{3\sqrt{\pi}\ln 2}n(s).
\end{equation}
Projecting now Eq. (\ref{Eq:URecSimple}) onto $n(s)$ (cf.~Sec.~\ref{SubSec:xiAv}) we get
\begin{equation}
c_{k-1}=c_k+\frac{1}{\sqrt{\pi}} \left(\frac{17+4\ln 2}{3\ln 2}\right)^2({\cal K}-k)
\end{equation}
implying
\begin{equation}
U_{\omega_1}(s)\sim \frac{1}{2\sqrt{\pi}} \left(\frac{17+4\ln 2}{3\ln 2}\right)^2{\cal K}^2 n(s).
\end{equation}
With the aid of Eq. (\ref{Eq:EaSFinal}) we  finally get
\begin{equation}
\left\langle\left(E_{\omega_1}^{\mathcal L}\right)^2\right\rangle \sim \frac{1}{4}\left(\frac{17+4\ln 2}{3\ln 2}\right)^2 {\cal K}^2=\frac{9}{4}\beta^2 {\cal K}^2. 
\label{Eq:EaSFinalFinal}
\end{equation}

\mysubsubsection{Averaging  $ E_{\omega_1}^{\mathcal L} E_{\omega_1}^{\mathcal R}$}
The average $\langle  E_{\omega_1}^{\mathcal L} E_{\omega_1}^{\mathcal R}\rangle$ can be presented as
\begin{equation}
\langle  E_{\omega_1}^{\mathcal L} E_{\omega_1}^{\mathcal R}\rangle=\frac{\sqrt{\pi}}{2}\int_0^\infty \tilde{R}^2_{\omega_1}(s).
\end{equation}
Using Eq. (\ref{Eq:Rass}) we find  
\begin{equation}
\langle  E_{\omega_1}^{\mathcal L} E_{\omega_1}^{\mathcal R}\rangle\sim \frac{\sqrt{\pi}}{2} \left(\frac{(17+4\ln 2){\cal K}}{3\sqrt{\pi}\ln 2}\right)^2\int_0^\infty ds n^2(s)= \frac{1}{4}\left(\frac{17+4\ln 2}{3\ln 2}\right)^2 {\cal K}^2=\frac{9}{4}\beta^2 {\cal K}^2.
\label{Eq:EREL}
\end{equation}

Equations (\ref{Eq:deltaxiBase}), (\ref{Eq:EaFinalFinal}), (\ref{Eq:EaSFinalFinal}) and (\ref{Eq:EREL}) immediately imply that the mesoscopic fluctuations scale linearly with $\log_2 L$, i.e.
\begin{equation}
\langle(\delta\overline{\zeta})^2\rangle\propto \log_2 L.
\end{equation}

\mysection{Non-zero top spin}
\label{Sec:S0}
To conclude this supplemental material, let us now show that our results on the asymptotic scaling of matrix elements, most notably Eq. (\ref{Eq:mainAns}), hold also 
outside the sector of the theory with $S_0=0$. 

Indeed, led us consider the sector with $S_0$ of the order of typical its typical value, $S_0\sim \sqrt{L}$. A straightforward generalization of the derivation presented in Sec. \ref{Sec:StatisticsOverBaseState} shows that in the asymptotic large-$L$ limit we still have (cf.~Eq.~(\ref{Eq:xibarFinal}))

\begin{equation}
\overline{\zeta}\approx \frac13\left( E^{\mathcal L}_{\omega_1}+ E^{\mathcal R}_{\omega_1}\right).
\end{equation}

Averaging with respect to the base state we then get (cf.~Eq.~(\ref{Eq:EaBasic1}))
\begin{eqnarray}
\langle \overline{\zeta}\rangle\approx \frac{2}{3}\langle E_{\omega_1}\rangle,\\
\langle E_{\omega_1}\rangle=\frac{1}{N_{\cal K}(S_0)}
\sum_{
\substack{S_1^{\mathcal L}, S_1^{\mathcal R}\\ 
|S_1^{\mathcal L}- S_1^{\mathcal R}| \leq S_0\leq S_1^{\mathcal L}+S_1^{\mathcal R}
}}^{L_{\cal K}/2} N_{{\cal K}-1}(S_1^{\mathcal R})R_{\omega_1}(S_1^{\mathcal L}).
\end{eqnarray}

In the continuum limit we thus get (with $s_0=S_0/\sqrt{L}$)
\begin{equation}
\langle E_{\omega_1}\rangle=\frac{e^{2s_0^2}}{2 \sqrt{2}s_0}\int_{|s_1-s_1^\prime|\leq \sqrt{2} s_0\leq s_1+s_1^\prime} d s_1 ds_1^\prime n(s_1^\prime)\tilde{R}_{\omega_1}(s_1).
\end{equation}
Also, using Eq. (\ref{Eq:Rass}) we find
\begin{equation}
\langle E_{\omega_1}\rangle=
-\frac{(17+4\ln 2){\cal K}}{3\sqrt{\pi}\ln 2}
\frac{e^{2s_0^2}}{2 \sqrt{2} s_0}\int_{|s_1-s_1^\prime|\leq \sqrt{2} s_0\leq s_1+s_1^\prime} d s_1 ds_1^\prime n(s_1^\prime)n(s_1)=-\frac{(17+4\ln 2)
}{6\ln 2} {\cal K}
\end{equation}
implying 
\begin{equation}
\langle\overline{\zeta}\rangle\sim -\beta \log_2 L\,, \qquad \beta=\frac{17+4\ln 2}{9\ln2}.
\end{equation}

In a similar manner one can show that the scaling laws for the the average  dispersion $\langle \sigma^2\rangle $ and mesoscopic fluctuations $\langle(\delta\overline{\zeta})^2\rangle$ remain unchanged in the case of arbitrary spin $S_0$

\begin{equation}
\langle \sigma^2\rangle\sim \langle\overline{\zeta}\rangle\sim  \log_2 L.
\end{equation}

\end{document}